\documentclass[12pt]{iopart}
\usepackage[pdftex]{graphicx}

\usepackage{amssymb}
\usepackage{mathabx}
\usepackage{iopams}
\usepackage{setstack}
\usepackage{rotating}

\begin{document}

\title[Distilling one, two and entangled pairs of photons from a quantum dot]{Distilling one, two and entangled pairs of photons from a quantum dot with cavity QED effects and spectral filtering}

\author{Elena del Valle} 

\address{Physik Department, Technische Universit{\"a}t M{\"u}nchen,
  James-Franck-Strasse, 85748 Garching, Germany}

\ead{elena.delvalle.reboul@gmail.com}

\begin{abstract}
  A quantum dot can be used as a source of one- and two-photon states
  and of polarisation entangled photon pairs. The emission of such
  states is investigated from the point of view of frequency-resolved
  two-photon correlations. These follow from a spectral filtering of
  the dot emission, which can be achieved either by using a cavity or
  by placing a number of interference filters before the
  detectors. The combination of these various options is used to
  iteratively refine the emission in a ``distillation'' process and
  arrive at highly correlated states with a high purity. So-called
  ``leapfrog processes'' where the system undergoes a direct
  transition from the biexciton state to the ground state by direct
  emission of two photons, are shown to be central to the quantum
  features of such sources. Optimum configurations are singled out in
  a global theoretical picture that unifies the various regimes of
  operation.
\end{abstract}

\newcommand{\ud}[1]{#1^{\dagger}} 
\newcommand{\bra}[1]{\left\langle #1\right|}
\newcommand{\ket}[1]{\left| #1\right\rangle}
\newcommand{\mean}[1]{\langle#1\rangle}

\maketitle

\section{Introduction}
\label{sec:1}

Quantum dots have proven in the recent years to be excellent platforms
for single photon sources~\cite{strauf07a,nguyen11a,matthiesen12a},
spin manipulation and coherent control at the
exciton~\cite{boyle09a,vamivakas09a,ramsay10a,simon11a,wu11a}, or
biexciton
level~\cite{chen02a,flissikowski04a,stufler06a,akimov06a,boyle10a,miyazawa10a},
or for entangled photon-pair
generation~\cite{akopian06a,stevenson06a,hafenbrak07a}. The
achievement of strong coupling between a single quantum dot and a
cavity mode~\cite{reithmaier04a,yoshie04a,peter05a} impulsed even
further these possibilities by increasing their efficiency and
output-collectability~\cite{dousse10a,hennessy07a} to the point of
reaching new regimes such as microlasing~\cite{nomura10a} or
two-photon emission~\cite{ota11a}.  The cavity mode can also serve as
a coupler between two distant dots~\cite{laucht10a,gallardo10a}.
Cavity QED effects are thus a powerful resource to exploit the quantum
features of a quantum
dot~\cite{schneebeli08a,kasprzak10a,arXiv_laussy11c,majumdar12b}.

There is an alternative way to control, engineer and purify the
emission of a quantum emitter which relies on extrinsic components at
the macroscopic level, in contrast with the intrinsic approach at the
microscopic level that supplements the quantum dot with a built-in
microcavity. Namely, one can use spectral filtering. This approach is
``extrinsic'' in the sense that the filters are placed between the
system which emits the light and the observer who detects it. As such,
it belongs more properly with the detection part. The filter can in
fact be modelling the finite resolution of a detector that is sensible
only within a given frequency window.
% This remains, however, of a fundamental character. 
% ~\cite{arnoldus84a,knoll84a,knoll86a,knoll86b,cresser87a}
% While the detection process is secondary in single-photon
% observables---such as in the photoluminescence, reflection and
% transmission spectra, etc., where the detection merely broadens and
% blurs a result that can be theoretically obtained in the
% ideal-detector limit---it turns out to be a necessary component to
% obtain physical observables when addressing multiple-photon
% correlations (we will consider here up to two-photon
% correlations). 
% This shows that one goes
% one step closer to the quantum features of the system when tracking
% its photon correlations, as it becomes mandatory to take into account
% the celebrated measurement problem.
%
In this text, to keep the discussion as simple as possible, we will
assume perfect detectors and describe the detection process through
spectral filters (this means that the detector has a better resolution
than the one imparted by the filter in front of it). Each filter is
theoretically fully specified by its frequency of detection and
linewidth. We will assume Lorentzian spectral shapes, which
corresponds to the case of most interference filters. Commonly used
spectral filter of this type are the thin-film filters and Fabry-Perot
interferometers (in the figures, we will sketch such filters as
dichroic bandpass filters, with different colours to imply different
frequencies.) Since they rely on interference effects, they are
basically cavities in weak coupling. This reinforces the main theme of
this text which is to investigate cavity effects on a quantum
emitter. The cavity itself can be, again, intrinsically part of the
heterostructure itself, all packaged on-chip, or extrinsically due to
the external filters. Combining these features, such as, filtering the
emission of a cavity-QED system, we arrive to the notion of
``distillation'' where the emitter sees its output increasingly
filtered by consecutive sequences to finally deliver a highly
correlated quantum state of high purity.

\begin{figure}[t] 
  \includegraphics[width=\linewidth]{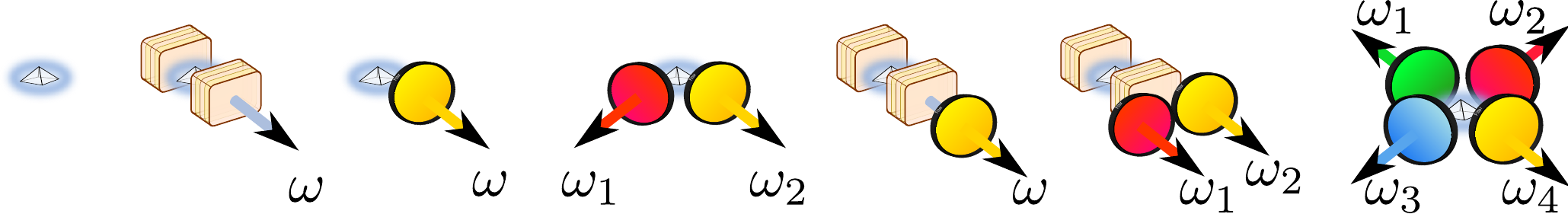}
  \caption{Sketch of the various schemes investigated in this text to
    study two-photon correlations from the light emitted by a quantum
    dot (left).  The various detection configurations are, from left
    to right: 1, the direct emission of the dot
    ($g^{(2)}[\sigma_\mathrm{H}]$), 2,~the enhanced and filtered
    emission of the dot by a cavity mode ($g^{(2)}[a]$); 3,~the
    filtered emission of the dot
    ($g_\Gamma^{(2)}[\sigma_\mathrm{H}](\omega;\omega)$), 4,~the
    two-photon spectroscopy of the dot
    ($g_\Gamma^{(2)}[\sigma_\mathrm{H}](\omega_1;\omega_2)$), 5,~the
    filtered emission of the dot-in-a-cavity emission
    ($g_\Gamma^{(2)}[a](\omega;\omega)$), 6,~the two-photon
    spectroscopy of the dot-in-a-cavity
    ($g_\Gamma^{(2)}[a](\omega_1;\omega_2)$) and 7,~the tomographic
    reconstruction of the density matrix for the
    polarisation-entangled photon pairs,
    $\theta_\Gamma(\omega_1,\omega_2;\omega_3,\omega_4)$.}
  \label{fig:1}
\end{figure}

While the idea is general and could be applied to a wealth of quantum
emitters, we concentrate here on a single quantum dot, sketched as a
little radiating pyramid in Fig.~\ref{fig:1}. Theoretically, it will
be described as a combination of two two-level systems, representing
two excitons of opposite spins.  Such a system will be used for the
generation of photons one by one or in pairs, with various types of
quantum correlations. The four-level system formed by the two possible
excitonic states (corresponding to orthogonal polarisations) and the
doubly occupied state, the biexciton, is ideal to switch from one type
of device to the other by simply selecting and enhancing the emission
at the different intrinsic
resonances~\cite{delvalle10a,delvalle11d,delvalle12b}. Figure~\ref{fig:1}
gives a summary of the various filtering and detection schemes that
will be applied, with the ``naked'' dot on the left. Its emission will
be considered both from within or without a cavity, with various
numbers of filters interceding. We will assume the microcavity both in
the weak and strong-coupling regimes.  The latter system has been
extensively
studied~\cite{delvalle10a,delvalle11d,delvalle12b,stace03a,schumacher12a}
and will be revisited here in the light of its spectral
filtering~\cite{delvalle12a} and distillation.

The rest of the text is organised as follows. In Sec.~\ref{sec:2}, we
present the system and its basic properties and we introduce the
\emph{two-photon spectrum} which is the counterpart at the two-photon
level of the photoluminescence spectrum at the single-photon one.  In
Sec.~\ref{sec:3} we provide the first application of two-photon
distillation, achieved via a cavity mode weakly coupled to the dot
transitions or through spectral filtering. In Sec.~\ref{sec:4}, we
compare the cavity filtering in the weak coupling regime with the
enhancement of the emission in the strong coupling regime. In
Sec.~\ref{sec:5}, we go one step further in the \emph{distillation} of
the two-photon emission and filter it from the cavity emission as
well. In Sec.~\ref{sec:6}, we consider one of the most popular
applications of the biexciton structure, the generation of
polarisation entangled photon pairs. In Sec.~\ref{sec:con} we draw some
conclusions.

\section{Two-photon spectrum from the quantum dot direct emission}
\label{sec:2}

\begin{figure}[t]
  \includegraphics[width=\linewidth]{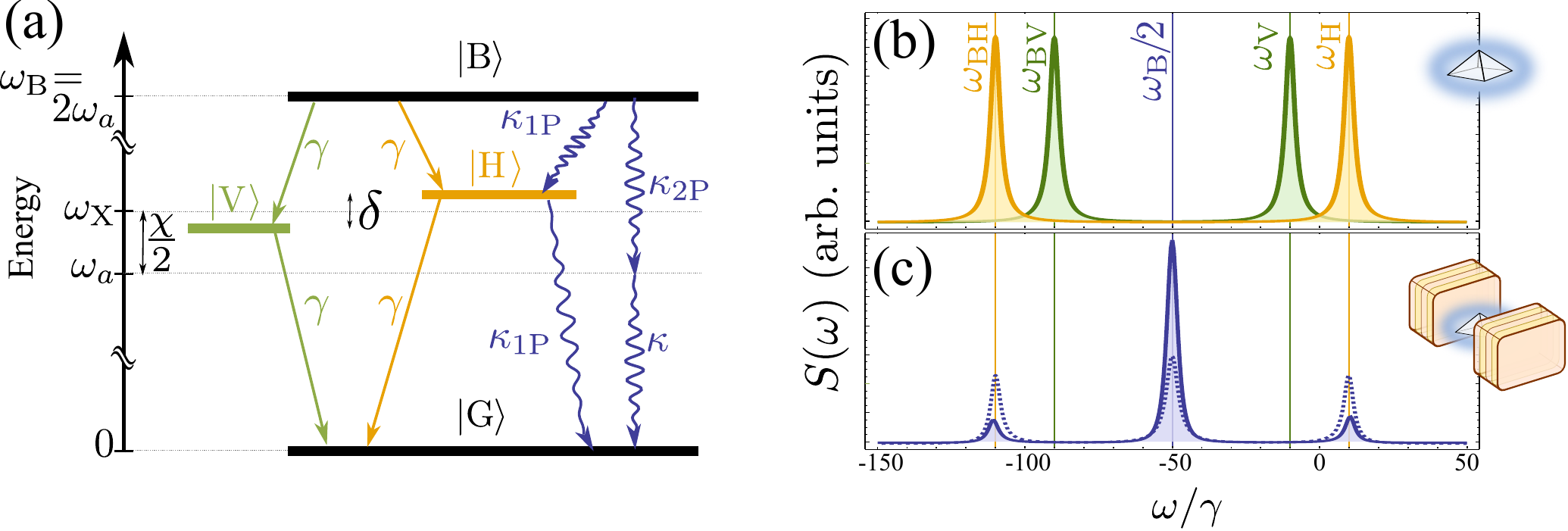}
  \caption{(a) Level scheme of the quantum dot investigated, modelled
    as a system able to accommodate two excitons $\ket{\mathrm{V}}$
    and $\ket{\mathrm{H}}$ in the linear polarisation basis, with an
    energy splitting $\delta$ between them and which, when present
    jointly, form a biexciton $\ket{\mathrm{B}}$ with binding
    energy~$\chi$. The excitons decay radiatively at a rate $\gamma$.
    When placed in a cavity (with linear polarisation~H), two extra
    decay channels are opened for the H-polarisation: through the
    one-photon cascade at rate $\kappa_\mathrm{1P}$, and the
    two-photon emission at rate $\kappa_\mathrm{2P}$. In the sketch,
    the cavity is placed at the two-photon resonance
    $\omega_a=\omega_\mathrm{B}/2$.  (b) PL spectra from the quantum
    dot system in H-polarisation (orange) consisting of two peaks at
    $\omega_\mathrm{BH}$ and $\omega_\mathrm{H}$, and in
    V-polarisation (green) consisting of two peaks at
    $\omega_\mathrm{BV}$ and $\omega_\mathrm{V}$.  (c) PL spectrum
    when the dot is placed inside a cavity, both for the case of
    strong ($g=\kappa$, solid line) and weak ($g\rightarrow 0$, dotted
    line) coupling.  A new peak appears at the centre from the
    two-photon emission. Parameters: $P=\gamma$, $\chi=100\gamma$,
    $\delta=20\gamma$, $\kappa=5\gamma$. We consider
    $\omega_\mathrm{X}\rightarrow 0$ as the reference frequency.}
  \label{fig:2}
\end{figure}

The system under analysis consists of a quantum dot that can host up
to two excitons with opposite spins. The corresponding orthogonal
basis of linear polarisations, Horizontal (H) and Vertical (V), reads
$\{\ket{\mathrm{G}},\ket{\mathrm{H}},\ket{\mathrm{V}},\ket{\mathrm{B}}\}$,
where G stands for the ground state, H and V for the single exciton
states and B for the biexciton or doubly occupied state. The four
level scheme that they form is depicted in Fig.~\ref{fig:2}(a). The
Hamiltonian of the system reads ($\hbar=1$):
\begin{equation}
  \label{eq:ThuApr14002810CEST2011}
  H_\mathrm{dot}=\Big(\omega_\mathrm{X}+\frac{\delta}{2}\Big)\ket{\mathrm{H}}\bra{\mathrm{H}}+\Big(\omega_\mathrm{X}-\frac{\delta}{2}\Big)\ket{\mathrm{V}}\bra{\mathrm{V}}+\omega_\mathrm{B}\ket{\mathrm{B}}\bra{\mathrm{B}}\,,
\end{equation}
where we allow the excitonic states to be split by a small
energy~$\delta$, as is typically the case experimentally, by the
so-called \emph{fine structure splitting}~\cite{ota11a}. The biexciton
at $\omega_\mathrm{B}=2\omega_\mathrm{X}-\chi$ is far detuned from
twice the exciton thanks to the binding energy, $\chi$, typically the
largest parameter in the system. We include the dot losses, at a rate
$\gamma$, and an incoherent continuous excitation (off-resonant
driving of the wetting layer), at a rate $P$, in both polarisations
$x=$H, V, in a master equation:
\begin{equation}
  \label{eq:TueDec20144239CET2011}
  \partial_t \rho= i[\rho,
  H_\mathrm{dot}]+\sum_{x\mathrm{=H,V}}\Big[\frac{\gamma}{2}\Big( \mathcal{L}_{\ket{\mathrm{G}}\bra{x}}+\mathcal{L}_{\ket{x}\bra{\mathrm{B}}}\Big)+\frac{P}{2}\Big(\mathcal{L}_{\ket{x}\bra{\mathrm{G}}}+\mathcal{L}_{\ket{\mathrm{B}}\bra{x}}\Big)\Big](\rho)\,,
\end{equation}
where~$\mathcal{L}_c(\rho)=2c\rho\ud{c}-\ud{c}c\rho-\rho \ud{c}c$ is
in the Lindblad form.

We assume in what follows an experimentally relevant situation,
$\chi=100\gamma$, $\delta=20\gamma$, and study the steady state under
$P=\gamma$, in which case all levels are equally populated (the
populations read $\rho_\mathrm{G}=\gamma^2/(P+\gamma)^2$,
$\rho_\mathrm{H}=\rho_\mathrm{V}=P\gamma/(P+\gamma)^2$ and
$\rho_\mathrm{B}=P^2/(P+\gamma)^2$). The photoluminescence spectra of
the system, $S(\omega)$, are shown in Fig.~\ref{fig:2}(b) for the H-
and V-polarised emission with orange and green lines respectively. The
four peaks are well separated thanks to the binding energy and fine
structure splitting, corresponding to the four transitions depicted in
panel (a) with the same colour code:
\begin{eqnarray}
  \label{eq:1}
  &\omega_\mathrm{BH}=\omega_\mathrm{B}-\omega_\mathrm{H}=\omega_\mathrm{X}-\chi-\delta/2=-110\gamma\,,\\
  &\omega_\mathrm{H}=\omega_\mathrm{X}+\delta/2=10\gamma\,,\\
  &\omega_\mathrm{BV}=\omega_\mathrm{B}-\omega_\mathrm{V}=\omega_\mathrm{X}-\chi+\delta/2=-90\gamma\,,\\
  &\omega_\mathrm{V}=\omega_\mathrm{X}-\delta/2=-10\gamma\,,
\end{eqnarray}
with $\omega_{X}\rightarrow 0$ as the reference and
FWHM~$\gamma_\mathrm{BH}=\gamma_\mathrm{BV}=3\gamma+P$,
$\gamma_\mathrm{H}=\gamma_\mathrm{V} =3 P+\gamma$. We concentrate on
the H-mode emission because it has the largest peak separation and
allows for the best filtering but all results apply similarly to the V
polarisation. The emission structure at the single-photon level is
very simple: two Lorentzian peaks are observed corresponding to the
upper and lower transitions,
\begin{equation}
  \label{eq:ThuNov15152435CET2012}
  S(\omega)=\frac{1}{\pi}\Big[ \rho_\mathrm{B}\frac{\gamma_\mathrm{BH}/2}{(\gamma_\mathrm{BH}/2)^2+(\omega-\omega_\mathrm{BH})^2}+\rho_\mathrm{H}\frac{\gamma_\mathrm{H}/2}{(\gamma_\mathrm{H}/2)^2+(\omega-\omega_\mathrm{H})^2}\Big]\,.
\end{equation}

The second-order coherence function of the H-emission in the steady
state reads:
\begin{equation}
  \label{eq:SunOct14202053CEST2012}
  g^{(2)}[\sigma_\mathrm{H}](\tau)=\mean{\sigma_\mathrm{H}^+(0)\sigma_\mathrm{H}^+
    (\tau)\sigma_\mathrm{H}(\tau)\sigma_\mathrm{H}(0)}/\mean{\sigma_\mathrm{H}^+\sigma_\mathrm{H}}^2
\end{equation}
where the H-photon destruction operator is defined as:
\begin{equation}
  \label{eq:SunOct14202018CEST2012}
  \sigma_\mathrm{H}=s_1+s_2\,,\quad\mathrm{with}\quad
  s_1=\ket{\mathrm{H}}\bra{\mathrm{B}} \quad\mathrm{and}\quad s_2=\ket{\mathrm{G}}\bra{\mathrm{H}}\,.
\end{equation}
In Eq.~(\ref{eq:SunOct14202053CEST2012}), we have specified the
channel of emission in square brackets since this will be an important
attribute in the rest of the text. The quantum-dot described with the
spin degree of freedom, exhibits uncorrelated statistics in the linear
polarisation:
\begin{equation}
  \label{eq:SunOct14154519CEST2012}
  g^{(2)}[\sigma_\mathrm{H}](\tau)=1\,.
\end{equation}
One recovers the expected antibunching of a two-level
system~\cite{mollow69a}, by turning to the intrinsic two-level systems
composing the quantum dot, namely, the spin-up and spin-down excitons:
$g^{(2)}[\sigma_\uparrow](0)=g^{(2)}[\sigma_\downarrow](0)=0$.  The
Pauli exclusion principle that holds for the spin
$\sigma_\updownarrows$, breaks in the linear polarisation, i.e., while
$\sigma_\updownarrows^2=0$, one has
$\sigma_\mathrm{H}^2=\ket{\mathrm{G}}\bra{\mathrm{B}}\neq0$. We can
find a simple explanation for this if we write the total correlations
in terms of the four contributions (different from zero):
\begin{equation}
  \label{eq:ThuNov15161649CET2012}
  \mean{\sigma_\mathrm{H}^+(0)\sigma_\mathrm{H}^+(\tau)\sigma_\mathrm{H}(\tau)\sigma_\mathrm{H}(0)}=\sum_{i,j=1,2}\mean{s_i^+(0) s_j^+(\tau) s_j(\tau)
  s_i(0)}\,,
\end{equation}
which are given by ($\tau\geq 0$):
\begin{eqnarray}
  \label{eq:ThuNov15154549CET2012}
  &g^{(2)}[s_1;s_1]
  (\tau)=(1-e^{-(\gamma+P)\tau})(1+\frac{\gamma}{P}e^{-(\gamma+P)\tau})\,,\nonumber  \\
  &g^{(2)}[s_2;s_2] (\tau)=(1-e^{-(\gamma+P)\tau})(1+\frac{P}{\gamma}e^{-(\gamma+P)\tau})\,,\nonumber \\
  &g^{(2)}[s_1;s_2] (\tau)=(1-e^{-(\gamma+P)\tau})+e^{-(\gamma+P)\tau}(2+\frac{\gamma}{P}+\frac{P}{\gamma})\,,\nonumber \\
  &g^{(2)}[s_2;s_1] (\tau)=(1-e^{-(\gamma+P)\tau})^2\,,
\end{eqnarray}
in their normalised form, $g^{(2)}[s_i;s_j](\tau)=\mean{s_i^+(0)
  s_j^+(\tau) s_j(\tau) s_i(0)} /(\mean{s_i^+s_i}\mean{s_j^+s_j})$. As
shown in Fig.~\ref{fig:3}(c) and (d) with pale grey lines, all these
functions are antibunched except for $g^{(2)}[s_1;s_2] (\tau)$, which
corresponds to the natural order in the H-cascaded emission of two
photons and is, consequently, bunched. It compensates fully the other
three terms, leading to total correlations of 1 for all $\tau$.

Fig.~\ref{fig:2} provides a clear picture o physical grounds of how
such a system can be used as a quantum emitter but it lacks even a
qualitative picture of how quantum correlations are
distributed. Fig.~\ref{fig:2}(b) merely shows where the system emits
light but nothing on how correlated is this emission. All these
crucial features are revealed in the \emph{two-photon spectrum}
$g_{\Gamma}^{(2)}(\omega_1;\omega_2,\tau)$~\cite{arXiv_tudela12a}. This
is the extension of the Glauber second order correlation
function---which quantifies the correlations between photons in their
arrival times---to frequency. By specifying both the energy and time
of arrivals of the photons, one provides an essentially complete
description of the system.  $\Gamma$ denotes the linewidth of the
frequency window of the filter over which this joint characterisation
is obtained. The corresponding time-resolution is given by its
inverse, $1/\Gamma$. It is a necessary variable without which
nonsensical or trivial results are obtained.

The two-photon spectrum unravels a large class of processes hidden in
single-photon spectroscopy and can be expected to become a standard
tool to characterise and engineer quantum sources. The computation of
such a quantity has remained a challenging task for theorists since
the mid-eighties~\cite{arnoldus84a,knoll86a,cresser87a}, until a
recent workaround~\cite{delvalle12a} has been found which allows
an exact numerical computation. It will be applied here for the first
time to the case of biexciton emission and used to understand,
characterize and enhance various processes useful for its quantum
emission. A detailed discussion on even more fundamental emitters is
given in Ref.~\cite{arXiv_tudela12a}.

% This method consists in coupling very weakly the system to two sensors
% or quantised modes, such that the dot dynamics remains unaltered, and
% looking into their intensity-intensity cross correlations such that
% . The sensors natural frequencies, $(\omega_1; \omega_2)$, decay
% rates, $\Gamma$, and the delay in the detection from each of them,
% $\tau$, can be varied to scan the relevant frequency and time ranges
% with the appropriate frequency and time resolution, given by $\Gamma$
% and $1/\Gamma$ respectively. In this particular case, we solve the
% coupling of two sensors with the H-mode only, obtaining
% $g_{\Gamma}^{(2)}[\sigma_\mathrm{H}](\omega_1;\omega_2,\tau)$.

Experimentally, the two-photon spectrum corresponds to the usual
Hanbury Brown--Twiss setup to measure second-order correlations through
photon counting, with filters or monochromators being placed in front
of the detectors to select two, in general different, frequency
windows. The technique has been amply used in the
laboratory~\cite{aspect80a,schrama91a,centenoneelen93a,moreau01a,akopian06a,stevenson06a,press07a,hennessy07a,kaniber08a,sallen10a,ulhaq12a}
but lacking hitherto a general theoretical description, the global
picture provided here has not yet been achieved experimentally. Note
finally that when considering correlations between equal frequencies,
$\omega_1=\omega_2$, the result is equivalent to placing a single
filter before measuring the correlations of the outcoming photon
stream~\cite{delvalle12a}.

Figure~\ref{fig:3}(a) shows the much richer landscape provided by the
frequency resolved second-order coherence function, that is, the
two-photon spectrum
$g^{(2)}_\Gamma[\sigma_\mathrm{H}](\omega_1;\omega_2,\tau)$, in
contrast with the one-photon spectrum $S(\omega)$ and the colour-blind
second order correlations $g^{(2)}[\sigma_\mathrm{H}]$,
Eq.~(\ref{eq:SunOct14154519CEST2012}).  It is shown at zero delay
($\tau=0$) with the sensor linewidths taken to
filter the full peaks ($\Gamma=5\gamma$). In such a case, one can see
well defined regions of enhancement and suppression of the
correlations: subpoissonian values ($<1$) are coloured in blue,
Poissonian ($=1$) in white and superpoissonian ($>1$) in red. This
figure is the backbone of this text. We now discuss in turns these
different regions where the quantum-dot operates as a quantum source
with different properties.

\section{Distilling single photons and photon pairs}
\label{sec:3}

\begin{figure}[t] 
  \includegraphics[width=\linewidth]{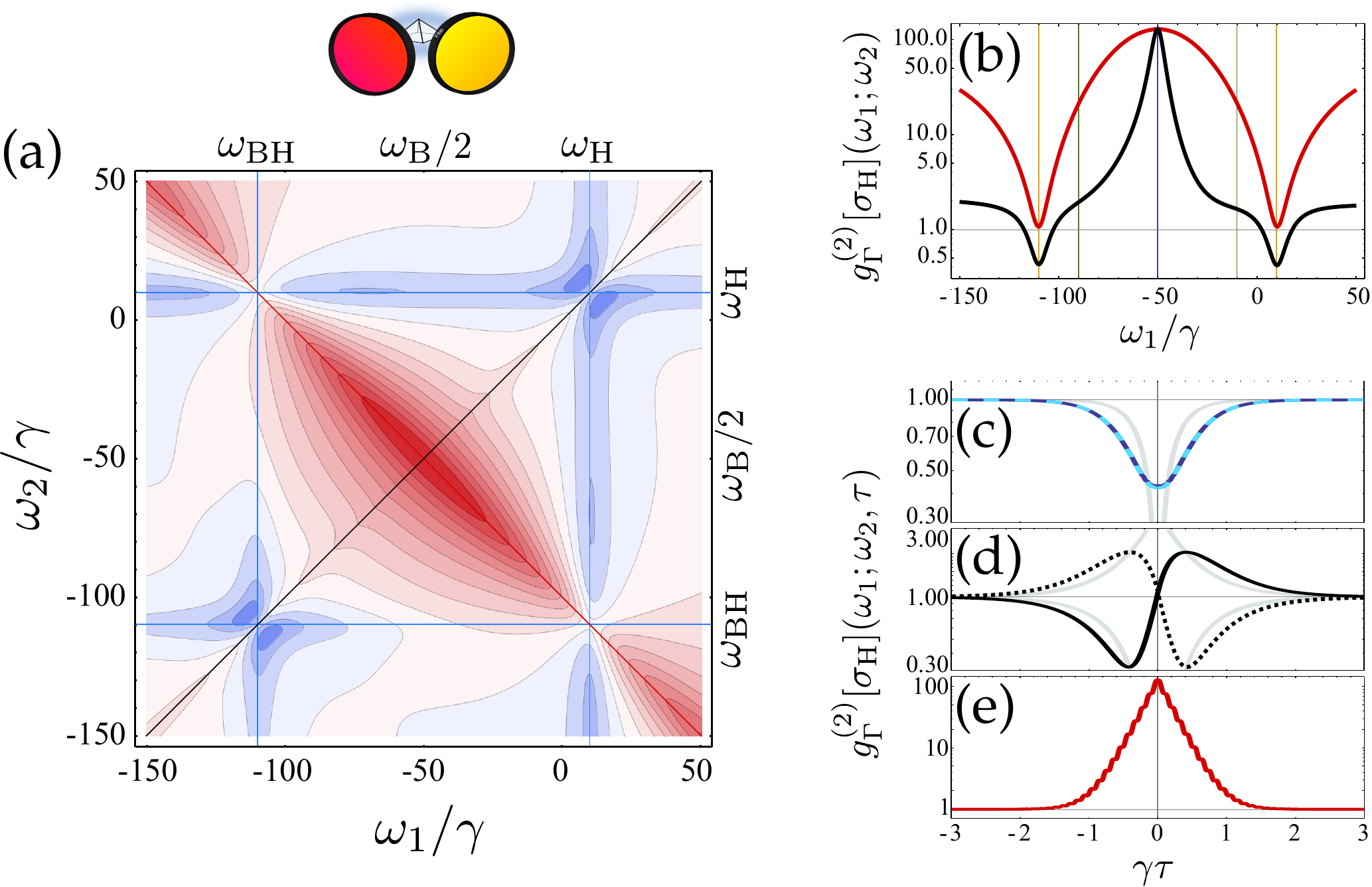}
  \caption{(a) Two-photon spectrum of a quantum dot (with a biexciton
    structure), with $\Gamma=5\gamma$. The density plot shows how the
    correlations between photons are distributed depending on their
    frequency of emission, from subpoissonian
    ($g_\Gamma^{2}[\sigma_\mathrm{H}](\omega_1;\omega_2)<1$, in blue)
    to superpoissonian ($>1$, in red) passing by Poissonian ($=1$, in
    white). The blue ``butterflies'' on the diagonal are typical of
    two-level systems. The antidiagonal corresponds to leapfrog
    processes with direct emission of two photons through an
    intermediate virtual state.  (b) Cuts from the density plot along
    the diagonal $\omega_2=\omega_1$ (in black) and the antidiagonal
    $\omega_2=-\chi-\omega_1$ (in red). The diagonal also corresponds
    to applying a single filter. (c--e) Comparison of the
    $\tau$-dynamics for three cases of interest: (c) bunching at the
    heart of the two blue butterflies on the diagonal at
    $(\omega_\mathrm{BH};\omega_\mathrm{BH})$ (solid dark blue) and
    $(\omega_\mathrm{H};\omega_\mathrm{H})$ (dashed clear blue), (d)
    cross-correlation of the peaks at
    $(\omega_\mathrm{BH};\omega_\mathrm{H})$ (solid) and vice-versa,
    $(\omega_\mathrm{H};\omega_\mathrm{BH})$ (dotted), showing the
    typical cascade behaviour and (e) the strong bunching at
    $(\omega_\mathrm{B}/2;\omega_\mathrm{b}/2)$ where two-photon
    emission is optimum. In (c) and (d) we also plot with pale grey
    lines the second order correlations of the effective corresponding
    operators, Eqs.~(\ref{eq:ThuNov15154549CET2012}). All scales are
    logarithmic. Parameters: $P=\gamma$, $\chi=100\gamma$,
    $\delta=20\gamma$.}
  \label{fig:3}
\end{figure}

\paragraph{Single-photon source:} When the filters are tuned to the
same frequency [diagonal black line in the $(\omega_1;\omega_2)$ space
in Fig.~\ref{fig:3}(a)], there is a systematic enhancement of the
bunching as compared to the surrounding regions due to the two
possibilities of detecting identical
photons~\cite{arXiv_tudela12a}. Despite this feature, that is
independent of the system dynamics, when both frequencies coincide
with one of the dot transitions,
$(\omega_\mathrm{H};\omega_\mathrm{H})$ or
$(\omega_\mathrm{BH};\omega_\mathrm{BH})$, there is a dip in the
correlations. This is more clearly shown in the cut at equal
frequencies, the black line in Fig.~\ref{fig:3}(b).  The blue
butterfly shape that is observed in the two-photon spectrum locally
around each of the dot transitions is characteristic of an isolated
two-level system~\cite{arXiv_tudela12a}. This zero delay information
is complemented by the antibunched $\tau$ dynamics, shown in
Fig.~\ref{fig:3}(c). The two dot resonances, upper and lower, coincide
in this case due to the symmetric conditions $P=\gamma$ but they are
typically different (the upper level being less antibunched at low
pump). Filtering and detection makes impossible to have a perfect
antibunching, getting closest to the ideal correlations
$g_\Gamma^{(2)}[\sigma_\mathrm{H}]
(\omega_\mathrm{H};\omega_\mathrm{H},\tau)\approx
g^{(2)}[s_1;s_1](\tau)$ from Eqs.~(\ref{eq:ThuNov15154549CET2012}), at
around $\Gamma\approx \chi/2$. At this point, the peaks are maximally
filtered with still negligible overlapping of the filters. It is
possible to derive a useful expression for the filtered correlations
at $\tau=0$, with $\Gamma\leq \chi/2$, in the limit:
\begin{equation}
  \label{eq:ThuNov15201716CET2012}
  \lim_{\chi\rightarrow \infty} g_\Gamma^{(2)}[\sigma_\mathrm{H}]  (\omega_\mathrm{H};\omega_\mathrm{H}) =\frac{2(P+\gamma)^2(3P+\gamma+\Gamma)(2\gamma+\Gamma)}{\gamma(P+\gamma+\Gamma)(2P+2\gamma+\Gamma)(3P+\gamma+3\Gamma)}\,,
\end{equation}
typically relevant in experiments.

All this shows that one can recover or optimise the quantum
features of a single photon source (antibunching) in a system whose
total emission is uncorrelated, by frequency filtering photons from
individual transitions. Consequently, the system should exhibit
two-photon blockade when probed by a resonant laser at frequency
$\omega_\mathrm{L}$ in resonance with the lower transition,
$\omega_\mathrm{L}=\omega_\mathrm{H}$. The antibunched emission of
each of the four filtered peaks of the spectrum, has been observed
experimentally~\cite{akopian06a}.

\paragraph{Cascaded two-photon emission:} When the filtering
frequencies match both the upper and lower quantum dot transitions,
i.e., $(\omega_\mathrm{BH};\omega_\mathrm{H})$, the correlations are
close to one at zero delay, like for the total emission
$g^{(2)}[\sigma_\mathrm{H}]$ (which is \emph{exactly} one). However,
although the latter is uncorrelated, since it remains equal to one at
all~$\tau$, the filtered cascade emission is \emph{not} uncorrelated
since it is close to unity only at zero delay and precisely because of
strong correlations that are, however, of an opposite nature at
positive and negative delays, i.e., showing enhancement for $\tau>0$
when photons are detected in the natural order that they are emitted,
and suppression for $\tau<0$ when the order is the opposite. This is
depicted in Fig.~\ref{fig:3}(d) where the solid line corresponds to
$(\omega_\mathrm{BH};\omega_\mathrm{H})$ and the dotted line to
exchanging the filters, $(\omega_\mathrm{H};\omega_\mathrm{BH})$. As
this also corresponds to detecting the photons in the opposite time
order, the two curves are exact mirror image of each other. 

The identification of the upper and lower transition photons with
frequency-blind operators [$g^{(2)}[s_1;s_2](\tau)$ in
Eqs.~(\ref{eq:ThuNov15154549CET2012})] provides crossed correlations
different to our exact and general frequency resolved functions,
specially at $\tau=0$, as shown in Fig.~\ref{fig:3}(d), where there is
a discontinuity for the approximated functions. The frequency resolved
functions have the typical smooth cascade shape that been observed
experimentally~\cite{moreau01a,akopian06a}. The dynamics at large
$\tau$, converges to the approximated functions only for
$\Gamma\approx \chi/2$.

\paragraph{Simultaneous two-photon emission:} For simultaneous
two-photon emission, the strongest feature lies on the antidiagonal
(red line) in Fig.~\ref{fig:3}(a), which is also shown as the solid
red line in Fig.~\ref{fig:3}(b). The strong bunching observed here,
when both frequencies are far from the system resonances
$\omega_\mathrm{BH}$ and $\omega_\mathrm{H}$, corresponds to a
two-photon deexcitation directly from the biexciton to the ground
state without passing by an intermediate real state. This two-photon
emission from a Hamiltonian, Eq.~(\ref{eq:ThuApr14002810CEST2011}),
that does not have a term to describe such a process is made possible
via a virtual state that arises in the quantum dynamics and that can
be revealed by the spectral filtering. As the intermediate virtual
state has no fixed energy and only the total energy
$\omega_1+\omega_2=\omega_\mathrm{B}$ needs be conserved, the
simultaneous two-photon emission is observed on the entire
antidiagonal (except, again, when touching a resonance, in which case
the cascade through real states takes over).  We call such processes
``\emph{leapfrog}'' as they jump over the intermediate excitonic
state~\cite{arXiv_tudela12a}. The largest bunching is found at the
central point, $\omega_1=\omega_2=\omega_\mathrm{B}/2=-\chi/2$, and at
the far-ends $\omega_1\ll\omega_\mathrm{BH}$ and
$\omega_1\gg\omega_\mathrm{H}$. Among them, the optimal point is that
where also the intensity of the two-photon emission is strong. The
frequency resolved Mandel $Q$ parameter takes into account both
correlations and the strength of the filtered signal~\cite{bel09a}:
\begin{equation}
  \label{eq:ThuOct18200154CEST2012}
  Q_\Gamma(\omega_1;\omega_2)=\sqrt{S_\Gamma(\omega_1)S_\Gamma(\omega_2)}\Big[g^{(2)}_\Gamma(\omega_1;\omega_2)-1\Big]  
\end{equation}
(where also for the single-photon spectra, the detection linewidth,
$\Gamma$, is taken into account~\cite{eberly77a}). As expected,
$Q_\Gamma[\sigma_\mathrm{H}](\omega_1;\omega_2)$ becomes negligible at
very large frequencies, far from the resonances of the system, and
reaches its maximum at the two-photon resonance,
$(\omega_\mathrm{B}/2;\omega_\mathrm{B}/2)$ (not shown). This latter
configuration is therefore the best candidate for the simultaneous
and, additionally, indistinguishable, emission of two photons.  The
bunching is shown in Fig.~\ref{fig:3}(e). The small and fast
oscillations are due to the effect of one-photon dynamics with the
real states but are unimportant for our discussion and would be
difficult to resolve experimentally. While the bunching in such a
configuration has not yet been observed experimentally, recently, Ota
\emph{et al.} successfully filtered the two-photon emission from the
biexciton with a cavity mode~\cite{ota11a}, which corroborates the
above discussion.

\section{Filtering and enhancing photon-pair emission from the quantum dot via a cavity mode}
\label{sec:4}

Large two-photon correlations are the starting point to create a
two-photon emission device. When they have been identified, the next
step is to increase their efficiency by enhancing the emission at the
right operational frequency. The typical way is to Purcell enhance the
emission through a cavity mode with the adequate polarisation and
strongly coupled with the dot transitions (at
resonance). Theoretically this amounts to adding to the master
equation~(\ref{eq:TueDec20144239CET2011}) a Hamiltonian part, that
accounts for the free cavity mode ($\omega_a$) and the coupling to the
dot (with strength $g$),
\begin{equation}
  \label{eq:ThuApr14002810CEST2011b}
  H_\mathrm{cav}=\omega_a\ud{a}a+g(\ud{a}\sigma_\mathrm{H}+a\ud{\sigma_\mathrm{H}})\,,
\end{equation}
along with a Lindblad term $\frac{\kappa}{2}\mathcal{L}_a( \rho)$,
that accounts for the cavity decay (at rate $\kappa$).

By placing the cavity mode at the two-photon resonance,
$\omega_a=\omega_\mathrm{B}/2$, the virtual leapfrog process becomes
real as it finds a real intermediate state in the form of a cavity
photon. The deexcitation of the biexciton to ground state is thereby
enhanced at a rate $\kappa_\mathrm{2PR}\approx (4g^2/\chi)^2/\kappa$,
producing the emission of two simultaneous and indistinguishable
cavity photons at this
frequency~\cite{delvalle10a,delvalle11d,delvalle12b}. There is as well
some probability that the cavity mediated deexcitation occurs in two
steps, through two different cavity photons at frequencies
$\omega_\mathrm{BH}$ and $\omega_\mathrm{H}$, at the same
rate~$\kappa_\mathrm{1PR}\approx 4g^2 \kappa/\chi^2$. The two
alternative paths are schematically depicted with curly blue arrows in
Fig.~\ref{fig:2}(a). The cavity being far from resonance with the dot
transitions, the ratio of two- versus one-photon emission can be
controlled by an appropriate choice of
parameters~\cite{delvalle11d}. We set $g=\kappa =5\gamma$, to be in
strong coupling regime and have
$\kappa_\mathrm{2P}=0.2\gamma>\kappa_\mathrm{1P}=0.05\gamma$, but with
a coupling weak enough for the system to emit cavity photons
efficiently. The cavity parameters are such that $\kappa >2P$ and the
pump does not disrupt the two-photon dynamics~\cite{delvalle12b}.  The
two-photon emission indeed dominates over the one-photon emission as
seen in Fig.~\ref{fig:2}(c), where the cavity spectrum is plotted with
a solid line: the central peak, corresponding to the simultaneous
two-photon emission, is more intense than the side peaks produced by
single photons. A better cavity (smaller $\kappa$) does not emit the
biexciton photons right away outside of the system, but spoils the
original (leapfrog) correlations and leads to smaller correlations in
the cavity emission $g^{(2)}[a]=\mean{a^\dagger a^\dagger
  aa}/\mean{a^\dagger a}^2\rightarrow 2$. This is shown in
Fig.~\ref{fig:4}(a) with a blue solid line. Our previous choice
$\kappa=5\gamma$, is close to that which maximises bunching (vertical
line in Fig.~\ref{fig:4}(a)).  A weak coupling due to small coupling
strength ($g\rightarrow0$), plotted with a blue dotted line, recovers
the case of a filter, discussed in the previous Section. The cavity
spectrum in this case, plotted with a blue dotted line in
Fig.~\ref{fig:2}(c), is no longer dominated by the two-photon
emission.

Regardless of the coupling strength $g$, the system goes into
weak-coupling at large enough $\kappa$, so both blue lines, solid and
dotted, converge to the same curve at $\kappa\rightarrow\infty$. The
cavity then filters the whole dot emission and recovers the total dot
correlations for the H-mode, $g^{(2)}[a]\rightarrow
g^{(2)}[\sigma_\mathrm{H}]=1$.  Note that while the bunching in
$g^{(2)}[a]$ is better in weak-coupling or with a filter, this is at
the price of decreasing the enhancement of the emission and,
therefore, the efficiency of the quantum device, as the total Mandel
$Q[a]$ parameter shows in Fig.~\ref{fig:4}(b).

\subsection{Distilling the two-photon emission from the cavity field}
\label{sec:5}

\begin{figure}[t] 
  \includegraphics[width=\linewidth]{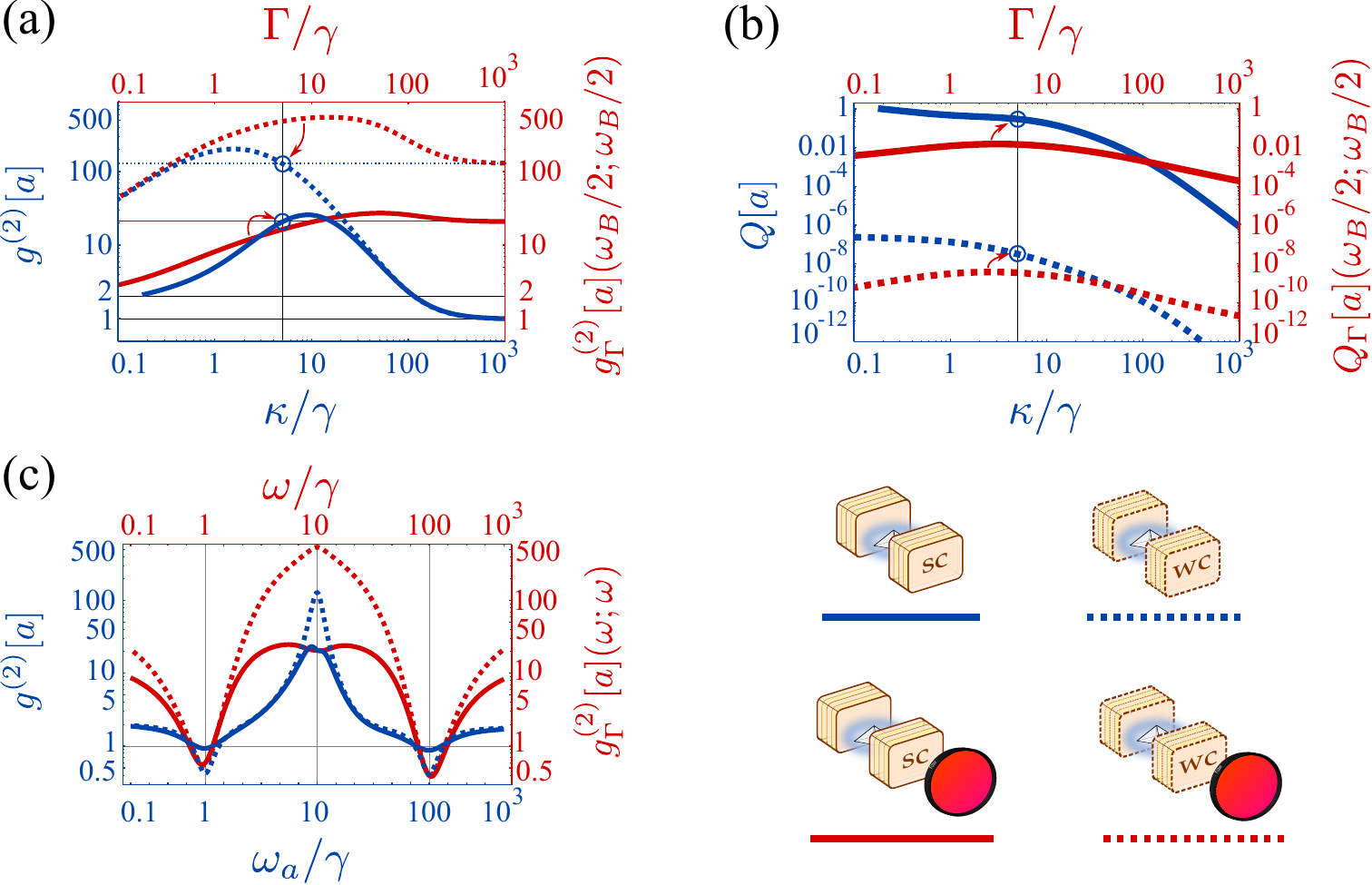}
  \caption{(a) Second order correlations of a cavity mode embedding a
    quantum dot in weak (dotted) and strong (solid) coupling.  Both
    the full, colour-blind, cavity emission (in blue) $g^{(2)}[a]$ and
    the frequency-resolved correlations at the two-photon resonance
    (in red)
    $g^{(2)}_\Gamma[a](\omega_\mathrm{B}/2;\omega_\mathrm{B}/2)$ are
    considered, with the former relating to the bottom axis (the
    cavity linewidth $\kappa$) and the latter to the upper axis (the
    detector linewidth $\Gamma$) at $\kappa=5\gamma$ (indicated by a
    circle). (b) Mandel $Q$ parameter in the same configuration as in
    (a), supplementing the information of the previous panel with the
    intensity of the emission. (c) Same as (a) but now as a function
    of the cavity frequency $\omega_a$ (bottom axis) for the
    colour-blind correlations (in blue) and of the detection frequency
    $\omega$ (upper axis) for the frequency-resolved correlations.
    Parameters: (a) the strong coupling is for $g=5\gamma$ and the
    weak coupling is in the limit of vanishing coupling $g\rightarrow
    0$ where the cavity is fully equivalent to a filter. Parameters:
    $P=\gamma$, $\chi=100\gamma$, $\delta=20\gamma$.}
  \label{fig:4}
\end{figure}

In view of the preceding results, we now consider the possibility to
further enhance the two-photon emission by filtering the cavity
photons from the central peak in Fig.~\ref{fig:2}(c). That is, we
study $g_\Gamma^{(2)}[a](\omega_1;\omega_2)$ for a cavity with
$\kappa=5\gamma$ and $\omega_a=\omega_\mathrm{B}/2$.  In this way,
first, the cavity acts as a filter, extracting the leapfrog emission
where it is most correlated, but also enhances specifically the
two-photon emission and, second, the filtering of the cavity emission
selects only those photons that truly come in pairs. Such a chain is
alike to a ``distillation'' process where the quantum emission is
successively refined.

The results are plotted with red lines in Fig.~\ref{fig:4}, for the
case $\kappa=5g$ pinpointed by circles on the blue lines. The filtered
cavity emission is indeed generally more strongly correlated at the
two-photon resonance than the unfiltered total cavity emission,
plotted in blue:
$g_\Gamma^{(2)}[a](\omega_\mathrm{B}/2;\omega_\mathrm{B}/2)\geq
g^{(2)}[a]$. This is so for all $\Gamma>\kappa$ for the cavity in
weak-coupling, where the distillation always enhances the
correlations. In strong coupling, the filter must strongly overlap
with the peak ($\Gamma\gg \kappa$). This is because the side peaks are
prominent in weak-coupling ($\kappa_\mathrm{2P}<\kappa_\mathrm{1P}$),
and the filtering efficiently suppress their detrimental effect,
whereas in strong-coupling, two-photon correlations are already close
to maximum thanks to the dominant central peak as seen in
Fig.~\ref{fig:2}(c), and it is therefore important for the filter to
strongly overlap with it. In all cases, at large enough $\Gamma$, the
full cavity correlations are recovered as expected:
$\lim_{\Gamma\rightarrow\infty}g_\Gamma^{(2)}[a](\omega_1;\omega_2)=g^{(2)}[a]$. In
Fig.~\ref{fig:4}(a), this means that the red lines converge to the
value projected by the circle on the blue lines at $\kappa=5\gamma$.
Here, again, filtering enhances correlations but reduces the number of
counts, as shown in Fig.~\ref{fig:4}(b).

In Fig.~\ref{fig:4}(c), we do the same analysis as in
Fig.~\ref{fig:3}(b) where there was no distillation.  We address the
same cases but now as a function of frequency, fixing the cavity decay
rate $\kappa=5\gamma$ and the filtering linewidth at
$\Gamma=10\gamma$.  In blue, we consider the cavity QED case in weak-
and strong-coupling, without filtering.  Since the weak-coupling limit
is identical to the single filter case, note that the blue dotted line
in Fig.~\ref{fig:4}(c) is identical to the black solid line in
Fig.~\ref{fig:3}(b). Off-resonance, the cavity acts as a simple filter
due to the reduction of the effective coupling.  The stronger coupling
to the cavity has an effect only when involving the real states, where
it spoils the correlations, less bunched at the two-photon resonance
$\omega_a=\omega_\mathrm{B}/2$, and less antibunched at the one-photon
resonances, $\omega_a=\omega_\mathrm{BH}$ or
$\omega_a=\omega_\mathrm{H}$. This shows again that useful quantum
correlations are obtained in a system where quantum processes are
Purcell enhanced and quickly transferred outside, rather than stored
and Rabi-cycled over within the cavity. The same is true for the red
line, further filtering the output. Finally, comparing solid lines
together, we see again that there is little if anything to be gained
by filtering in strong-coupling, whereas in weak-coupling, the
enhancement is considerable.  As a summary, the filtering of the
weakly coupled cavity (red dotted), provides the strongest
correlations (at the cost of the available intensity), corresponding
to distilling the photon pairs out of the original dot spectrum
without any additional enhancement.

\begin{figure}[t] 
 \includegraphics[width=0.75\linewidth]{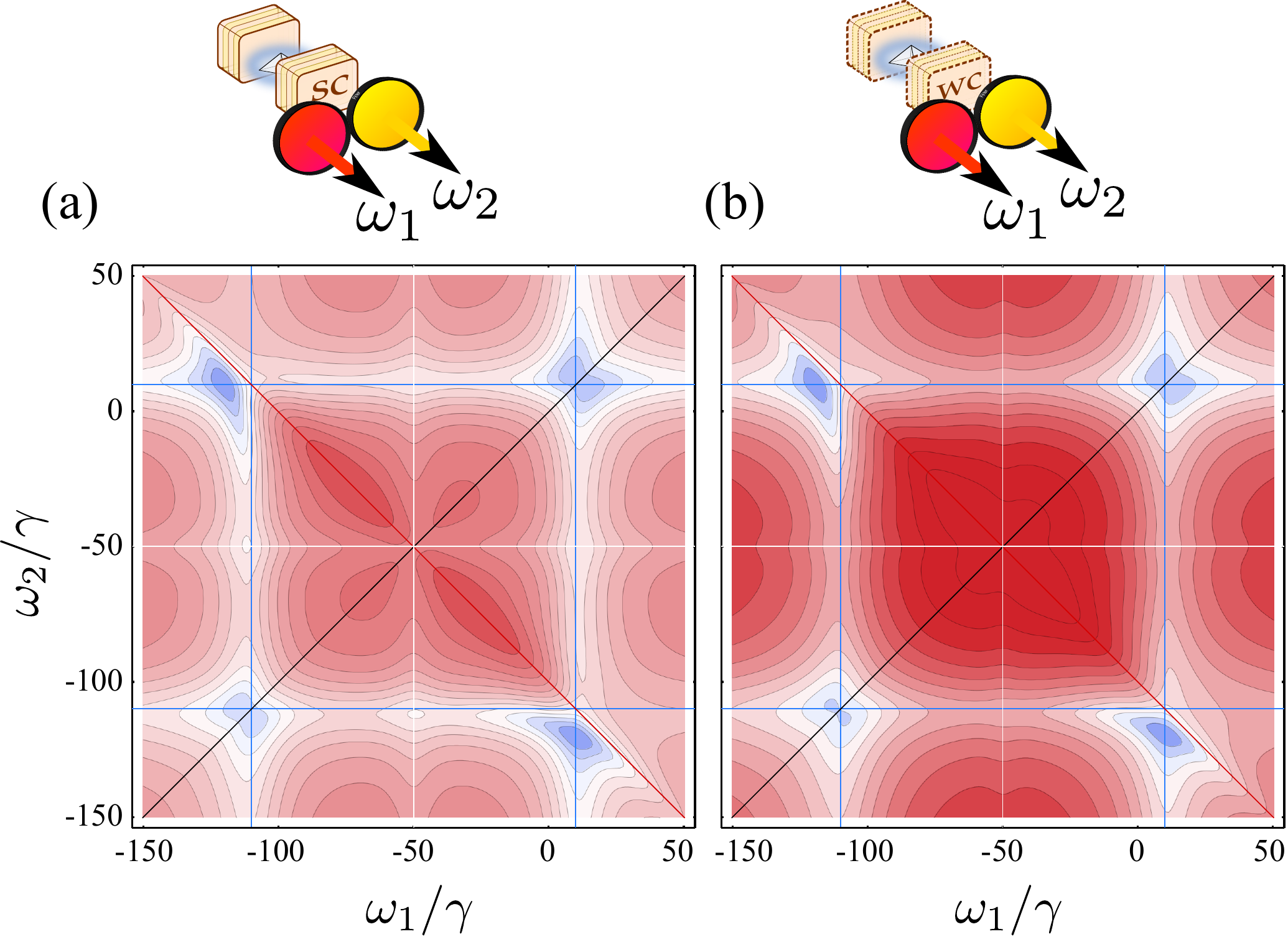}
 \caption{Two-photon spectra $g^{(2)}_\Gamma[a](\omega_1;\omega_2)$
   for a quantum dot in a cavity in the (a) strong, $g=\kappa$, and
   (b) weak coupling regime, when the cavity is at the two-photon
   resonance, $\omega_a=\omega_\mathrm{B}/2$. The bunching regions are
   strengthened by the cavity (the total, colour-blind correlations
   are $g^{(2)}[a]\approx 21$). A horizontal and vertical structure
   also emerges at the two-photon resonance, betraying the emergence
   of real states. These are stronger the stronger the coupling.  The
   same logarithmic scale as Fig.~\ref{fig:3}(a) applies, so all three
   figures can be compared directly.  Parameters: $\kappa=5\gamma$,
   $P=\gamma$, $\chi=100\gamma$, $\delta=20\gamma$ and
   $\Gamma=10\kappa$.}
  \label{fig:5}
\end{figure}

In Fig.~\ref{fig:5}, we show the full two-photon spectra for a quantum
dot in a cavity, in both (a) strong- and (b) weak-coupling.  The same
colour code and logarithmic scale is used as in Fig.~\ref{fig:3}(a),
for comparison. The antibunching regions on the diagonal and the
bunching ones on the antidiagonal are qualitatively similar to the
filtered dot emission, but antibunching is milder and less extended
while correlations are, respectively, weaker (stronger) at the central
point due to the saturated (efficient) distillation in strong (weak)
coupling. Another striking feature added by the cavity is the
appearance of an additional pattern of horizontal and vertical lines
at $\omega_a=\omega_\mathrm{B}/2$. While diagonal and anti-diagonal
features correspond to virtual processes, horizontal and vertical stem
from real processes, that pin the correlations at their own
frequency. Therefore, the new features are a further illustration that
the two-photon emission becomes a real resonance of the cavity-dot
system, in contrast with Fig.~\ref{fig:3}(a) where it was virtual. The
effect is more pronounced in the strong-coupling regime since this is
the case where the new state is better defined.  Another qualitative
difference between Fig.~\ref{fig:5} and Fig.~\ref{fig:3}(a) is in the
regions surrounding the cascade configuration,
$(\omega_\mathrm{BH};\omega_\mathrm{H})$, which has changed shape
around two antibunching spots. This is due to the fact that the
single-photon cascade is much less likely to happen through the cavity
mode than the direct deexcitation of the dot, as
$\kappa_\mathrm{1P}=0.05\gamma\ll \gamma$. Even if the first photon
from the biexciton emission decays through the cavity, the second will
most likely not. This new two antibunching spots are slightly pushed
to the left of the red line by the leapfrog bunching line and the
presence of the V-polarised resonances.

%\vfill\eject
\section{Distilling entangled photon pairs}
\label{sec:6}

One of the most sophisticated applications of the biexciton structure
in a quantum dot is as a source of polarisation entangled photon
pairs~\cite{benson00a,akopian06a,stevenson06a,troiani06a,meirom08a,pfanner08a,avron08a,johne08a,pathak09a,dousse10a,carmele10b,carmele11a,schumacher12a,poddubny12a}. Without
the fine-structure splitting, $\delta=0$, the two possible two-photon
deexcitation paths are indistinguishable except for their polarisation
degree of freedom (H or V), producing equal frequency photon pairs
$(\omega_\mathrm{BX};\omega_\mathrm{X})$, with
$\omega_\mathrm{BX}=\omega_\mathrm{B}-\omega_\mathrm{X}$. This results
in the polarisation entangled state:
\begin{equation}
  \label{eq:MonOct15142628CEST2012}
  \ket{\psi}=\frac{1}{\sqrt2}\left(\ket{\mathrm{H (\omega_\mathrm{BX}),H(\omega_\mathrm{X})}}+e^{-i\phi}\ket{\mathrm{V(\omega_\mathrm{BX}),V (\omega_\mathrm{X})}}\right)\,.
\end{equation}
The splitting $\delta$ provides ``which path''
information~\cite{scully91a} that spoils indistinguishability,
producing a state entangled in both frequency and
polarisation~\cite{stace03a}:
\begin{equation}
  \label{eq:MonOct15143347CEST2012}
  \ket{\psi'}=\frac{1}{\sqrt2}\left(\ket{\mathrm{H(\omega_\mathrm{BH}),H(\omega_\mathrm{H})}}+e^{-i\phi'}\ket{\mathrm{V(\omega_\mathrm{BV}),V(\omega_\mathrm{V})}}\right)\,.
\end{equation}
Although these doubly entangled states are useful for some quantum
applications~\cite{chuan09a,wang09a}, it is typically desirable to
erase the frequency information and recover polarisation-only
entangled pairs. Among other solutions, such as canceling the built-in
splitting externally~\cite{stevenson06a,trotta12a}, filtering has been
implemented with $\omega_1=\omega_\mathrm{BX}$ and
$\omega_2=\omega_\mathrm{X}$ to make the pairs identical in
frequencies again~\cite{stace03a,akopian06a,meirom08a}, at the cost of
increasing the randomness of the source (making it less
``on-demand''). Recently, the cavity filtering of the polarisation
entangled photon pairs with $\omega_a=\omega_\mathrm{B}/2$ has been
proposed by Schumacher \emph{et al.}~\cite{schumacher12a}, taking
advantage of the additional two-photon
enhancement~\cite{delvalle11d}. Let us revisit these effects in the
light of the previous results.

The properties of the output photons can be obtained from the
two-photon state density matrix, $\theta(\tau)$, reconstructed in the
basis
$\{\ket{\mathrm{H1,H2}},\ket{\mathrm{H1,V4}},\ket{\mathrm{V3,H2}},\ket{\mathrm{V3,V4}}\}$,
denoting by $\ket{xi}$ with $x=$H or V and $i=1, 2, 3, 4$, the state
$\ket{x(\omega_i)}$. The frequencies $\omega_i$ are, in general,
different. The second photon is detected with a delay $\tau$ with
respect to the first one (detected in the steady state). Let us
express this matrix in terms of frequency resolved correlators, as is
typically done in the
literature~\cite{troiani06a,pfanner08a}. However, in contrast to
previous approaches, we do not identify the photons with the
transition from which they may come from (using the dot operators
$\ket{\mathrm{G}}\bra{x}$, $\ket{x}\bra{\mathrm{B}}$ with, again, $x$
standing for either H or V) but with their measurable properties, that
is, polarisation, frequency and time of detection (for a given filter
window). This is a more accurate description of the experimental
situation where a given photon can come from any dot transition and
any transition can produce photons at any frequency and time with some
probability. We describe the experiments by considering four different
filters, that is, including all degrees of freedom of the emitted
photons in the description. Each detected filtered photon corresponds
to the application of the filter operator $\varsigma_j$ with $j=$H1,
H2, V3, V4 corresponding to its coupling to the H or V dot transitions
with $\omega_i$ frequencies. Then, the two-photon matrix
$\theta'_\Gamma(\tau)$ (the prime refers to the lack of normalisation)
corresponding to a tomographic measurement is theoretically
modelled~as:
\begin{eqnarray}
  \label{eq:SatFeb6235837WET2010}
  \fl\theta_\Gamma'(\tau)=
  \left(
    \begin{array}{cccc}
      \mean{n_\mathrm{H1}(0)n_\mathrm{H2}(\tau)}  & \mean{n_\mathrm{H1}(0) [\varsigma_\mathrm{H2}^+\varsigma_\mathrm{V4}](\tau)} &  \mean{[\varsigma_\mathrm{H1}^+\varsigma_\mathrm{V3}](0)n_\mathrm{H2}(\tau)}  \\
      \mathrm{h.c.} & \mean{n_\mathrm{H1}(0)n_\mathrm{V4}(\tau)} & \mean{[\varsigma_\mathrm{H1}^+\varsigma_\mathrm{V3}](0)  [\varsigma_\mathrm{V4}^+\varsigma_\mathrm{H2}] (\tau)} \\
      \mathrm{h.c.} & \mathrm{h.c.}  & \mean{n_\mathrm{V3}(0)n_\mathrm{H2}(\tau)} \\
      \mathrm{h.c.} & \mathrm{h.c.}  & \mathrm{h.c.}  
    \end{array}
    \right.\nonumber\\
    \hskip7.25cm\left.
      \begin{array}{c}
      \mean{[\varsigma_\mathrm{H1}^+\varsigma_\mathrm{V3}](0)  [\varsigma_\mathrm{H2}^+\varsigma_\mathrm{V4}] (\tau)}  \\
      \mean{[\varsigma_\mathrm{H1}^+\varsigma_\mathrm{V3}](0)  n_\mathrm{V4} (\tau)}  \\
      \mean{n_\mathrm{V3}(0)  [\varsigma_\mathrm{H2}^+\varsigma_\mathrm{V4}] (\tau)}  \\
      \mean{n_\mathrm{V3}(0)n_\mathrm{V4}(\tau)}
      \end{array}
    \right)\, ,
\end{eqnarray}
where $n_i=\varsigma_i^+ \varsigma_i$ [we have dropped the frequency
dependence in the notation, writing $\theta_\Gamma '(\tau)$ instead of
$\theta_\Gamma '(\omega_1,\omega_2;\omega_3,\omega_4,\tau)$]. Since a
weakly coupled cavity mode behaves as a filter, this tomographic
procedure is equivalent to considering the four dot transitions
coupled to four different cavity modes with the corresponding
polarisations, central frequencies and decay
rates~\cite{schumacher12a,poddubny12a}.  Unlike in other works where
for various reasons and particular cases, some of the elements in
$\theta_\Gamma '(\tau)$ are set to zero or considered equal, here we
keep the full matrix with no a~priori assumptions since, in general,
it may not reduce to a simpler form due to the incoherent pumping,
pure dephasing, frequency filtering and fine-structure splitting.

There are essentially two ways to quantify the degree of entanglement
from the density matrix~$\theta_\Gamma'(\tau)$.  The most
straightforward is to consider the $\tau$-dependent matrix directly,
which merely requires normalisation at each time~$\tau$, yielding
$\theta_\Gamma(\tau)=\theta_\Gamma '(\tau)/\mathrm{Tr}[\theta_\Gamma
'(\tau)]$. The physical interpretation is that of photon pairs emitted
with a delay $\tau$, that is to say, within the time-resolution
$1/\Gamma$ of the filter or cavity~\cite{stace03a,poddubny12a}. In
particular, the zero-delay matrix, $\theta_\Gamma(0)$, represents the
emission of two \emph{simultaneous}
photons~\cite{carmele10b,carmele11a}.  The second approach is closer
to the experimental measurement which averages over time. In this
case, one considers the integrated quantity
$\Theta_\Gamma(\tau_\mathrm{max})=\big(\int_0^{\tau_\mathrm{max}}
\theta_\Gamma '(\tau)\,d\tau\big)/\mathcal{N}$, that averages over all
possible emitted pairs from the
system~\cite{akopian06a,troiani06a}. It is also normalised (by
$\mathcal{N}$), but after integration, so that the two approaches are
not directly related to each other and present alternative aspects of
the problem, discussed in detail in the following.  Without the cutoff
delay $\tau_\mathrm{max}$, the integral diverges due to the continuous
pumping.

The degree of entanglement of any bipartite system represented by a
$4\times 4$ density matrix $\theta$, can be quantified by the
\emph{concurrence}, $C$, which ranges from~0 (separable states) to~1
(maximally entangled states)\footnote{Its definition reads
  $C\equiv[\mathrm{max}\{0,\sqrt{\lambda_1}-
  \sqrt{\lambda_2}-\sqrt{\lambda_3}- \sqrt{\lambda_4}\}]$, where
  $\{\lambda_1, \lambda_2, \lambda_3, \lambda_4 \}$ are the
  eigenvalues in decreasing order of the matrix $\theta T \theta^*T$,
  with $T$ an antidiagonal matrix with elements $\{
  -1,1,1,-1\}$.}~\cite{wootters98a}. High values of the concurrence
require high degrees of purity in the system~\cite{munro01a}, being,
for instance, impossible to extract any entanglement from a maximally
mixed state (in which case all the four basis states occur with the
same probability).  The filtered density matrices,
$\Theta_\Gamma(\tau_\mathrm{max})$ and $\theta_\Gamma(\tau)$, provide
each their own concurrence that we will denote
$C_\Gamma^\mathrm{int}(\tau_\mathrm{max})$ and $C_\Gamma(\tau)$,
respectively.  

\begin{figure}[!htbp] 
  \includegraphics[width=\linewidth]{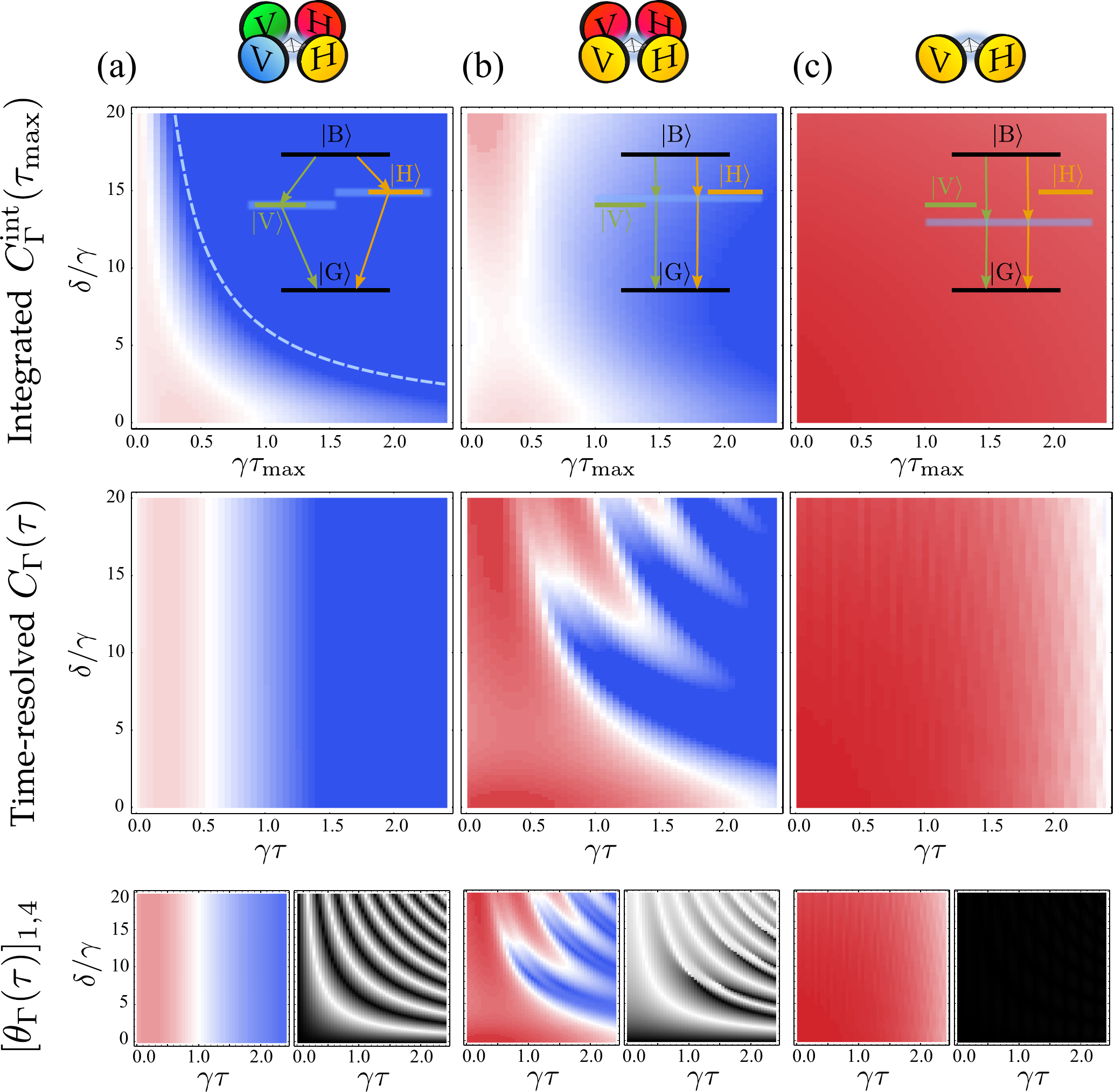}
  \caption{Concurrence for three different schemes of entanglement
    distillation (as sketched above each column): (a) filtering the
    four different dot resonances in their cascade through the real
    states, (b) filtering at two different frequencies
    $\omega_1=\omega_\mathrm{BX}$ and $\omega_2=\omega_\mathrm{X}$,
    degenerate in both polarisation decay paths, (c) filtering at the
    same frequency $\omega_\mathrm{B}/2$, i.e., at the two-photon
    resonance.  The upper panels show the time-integrated concurrence
    $C_\Gamma^\mathrm{int}(\tau_\mathrm{max})$ as a function of its
    cutoff $\tau_\mathrm{max}$ and the fine-structure splitting
    $\delta$.  In (a) the line $\tau_\mathrm{max}=2\pi/\delta$
    bounding the region with entanglement is superimposed. The
    intermediate panels show the instantaneous concurrence
    $C_\Gamma(\tau)$ as a function of delay and splitting. The lowest
    panels show the modulus and phase of the off-diagonal element
    $[\theta_\Gamma(\tau)]_{1,4}$ typically used in the literature to
    quantify entanglement. All these plots show that as far as the
    degree of entanglement is concerned, the leapfrog emission is the
    best configuration and is optimum at the two-photon resonance.
    The concurrence colour code is blue for 0, white for 0.5 and red
    for 1. In the modulus density plots, the colour code is blue for
    0, white for 0.25 and red for values $\geq 0.5$. In the phase
    density plots, the colour code is black for values
    approaching~$-\pi$ and white for $\pi$.  Parameters: $P=\gamma$,
    $\chi=100\gamma$, $\Gamma=2\gamma$.}
  \label{fig:6}
\end{figure}

We begin by considering the standard cascade configuration by
detecting photons at the dot resonances, i.e.,
$\omega_1=\omega_\mathrm{BH}$, $\omega_2=\omega_\mathrm{H}$,
$\omega_3=\omega_\mathrm{BV}$, $\omega_4=\omega_\mathrm{V}$, as
sketched in the inset of Fig.~\ref{fig:6}(a).  The upper density plot
shows $C_\Gamma^\mathrm{int}(\tau_\mathrm{max})$ and the density plot
below shows the time-resolved concurrence $C_\Gamma(\tau)$, as a
function of $\tau_\mathrm{max}$ or $\tau$ and $\delta$, for
$\Gamma=2\gamma$. The two concurrences,
$C_\Gamma^\mathrm{int}(\tau_\mathrm{max})$ and $C_\Gamma(\tau)$, are
qualitatively different except at $\tau=\tau_\mathrm{max}=0$ where
they are equal by definition. They also have in common that the
maximum concurrence is not achieved at zero delay [it is most visible
as the darker red area around
$(\gamma\tau_\mathrm{max},\delta/\gamma)\approx(0.4,0)$]. This is
because this filtering scheme relies on the real-states deexcitation
of the dot levels, and thus exhibits the typical delay from the
cascade-type dynamics of correlations [see Fig.~\ref{fig:3}(d)].  The
major departure between the two is that the decay of
$C^\mathrm{int}_\Gamma(\tau_\mathrm{max})$ is strongly dependent on
$\delta$, while $C_\Gamma(\tau)$ is not. With no splitting, at
$\delta=0$, the ideal symmetrical four-level structure efficiently
produces the entangled state $\ket{\psi}$ and the concurrence is
maximum both for the integral and the time-resolved forms.  The decay
time of $C_\Gamma(\tau)$ is the simplest to understand as, when
filtering full peaks, it is merely related to the reloading time of
the biexciton, of the order of the inverse pumping rate, $\sim
2/P$~\cite{poddubny12a}.  The asymmetry due to a nonzero splitting in
the four-level system causes an unbalanced dynamics of deexcitation
via the H and V polarisations. The entanglement in the form
$\ket{\psi}$ is downgraded to $\ket{\psi'}$. The fact that the
concurrence $C_\Gamma(\tau)$ is not affected shows that it accounts
for the degree of entanglement in both polarisation and frequency,
which is oblivious to the ``which path'' information that the last
variable provides. On the other hand,
$C_\Gamma^\mathrm{int}(\tau_\mathrm{max})$ is suppressed by the
splitting as fast as $\tau_\mathrm{max} >2\pi/\delta$ (this boundary
is superimposed to the density plot). That is, as soon as the
integration interval is large enough to resolve it. The integrated
$C_\Gamma^\mathrm{int}(\tau_\mathrm{max})$, therefore, accounts for
the degree of entanglement in polarisation only, and is destroyed by
the ``which path'' information provided by the different
frequencies. The exact mechanism at work to erase the entanglement is
shown in the lower row of Fig.~\ref{fig:6}, which displays the upper
right matrix element $[\theta_\Gamma(\tau)]_{1,4}$ of the density
matrix, that is the most responsible for purporting entanglement. The
modulus is shown on the left (with blue meaning 0 and red meaning
$\ge0.5$) and the phase on the right (in black and white). The
time-resolved concurrence $C_\Gamma(\tau)$ is very similar to the
modulus of $[\theta_\Gamma(\tau)]_{1,4}$ which justifies the
approximation often-made in the literature
$C_\Gamma(\tau)=2(|[\theta_\Gamma(\tau)]_{1,4}|-[\theta_\Gamma(\tau)]_{2,2})$
with $[\theta_\Gamma(\tau)]_{2,2}$ small~\cite{troiani06a}. Although
each photon pair is entangled, it is so in the state $\ket{\psi'}$
with a phase, $\phi'=-\pi +\delta \tau$, that accumulates with $\tau$
at a rate $\delta$~\cite{stace03a}, but this does not matter as far as
instantaneous entanglement is concerned, and this is why the splitting
does not affect $C_\Gamma(\tau)$. On the other hand, when integrating
over time, the varying phase, that completes a $2\pi$-cycle at
intervals $2\pi/\delta$, randomises the quantum superposition and
results in a classical mixture. This is why the splitting does
completely destroy $C_\Gamma^\mathrm{int}(\tau_\mathrm{max})$ for
$\tau_\mathrm{max} >2\pi/\delta$. And this is how the system restores
the ``which path'' information: given enough time, if the splitting is
large enough, the photons loose their quantum coherence due to the
averaging out of the relative phase between them because of their
distinguishable frequency.

Another possible configuration proposed in the
literature~\cite{stace03a,akopian06a,meirom08a} is sketched in the
inset of Fig.~\ref{fig:6}(b).  Photons are detected at the frequency
that lies between the two polarisations,
$\omega_1=\omega_3=\omega_\mathrm{BX}$,
$\omega_2=\omega_4=\omega_\mathrm{X}$. For small splittings, the
entanglement production scheme still relies on the cascade and real
deexcitation of the dot levels. Therefore, the behaviour of
$C_\Gamma^\mathrm{int}(\tau_\mathrm{max})$ is similar to (a) (decaying
with $\delta$) but slightly improved by the fact that the
contributions to the density matrix are more balanced by filtering
in-between the levels. For splittings large enough to allow the
formation of leapfrog emission in both paths, $\delta\gg \Gamma$,
there is a striking change of trend and
$C_\Gamma^\mathrm{int}(\tau_\mathrm{max})$ remains finite at longer
delays $\tau_\mathrm{max}$ when increasing $\delta$. This is a clear
sign that the entanglement relies on a different type of emission,
namely simultaneous leapfrog photon pairs, rather than a cascade
through real states. Accordingly, the intensity is reduced as compared
to that obtained at smaller $\delta$ or to the non-degenerate cascade
in (a), but the bunching is stronger, as evidenced by the two-photon
spectrum, and results in a larger degree of entanglement than at
$\delta=0$. A similar result is obtained qualitatively with the
time-resolved concurrence, a strong resurgence of entanglement with
$\delta$, indicating again very high correlations in this
configuration. Note finally that the phase becomes much more constant
with increasing $\delta$, resulting in the persistence of the
correlation in the time-integrated concurrence. This is because the
two-photon emission is through the leapfrog processes. Since the
latter, by definition, involve intermediate virtual states that are
degenerate in frequency, this is a built-in mechanism to suppress the
splitting and not suffer from the ``which path'' information as when
passing through the real states.

Finally, a configuration proposed more recently in the
literature~\cite{schumacher12a} is the two-photon resonance, with four
equal frequencies,
$\omega_1=\omega_2=\omega_3=\omega_4=\omega_\mathrm{B}/2$, as sketched
in the inset of Fig.~\ref{fig:6}(c). This provides a two-photon source
in both polarisations that can be enhanced via two cavity modes with
orthogonal polarisations~\cite{delvalle11d}. Remarkably, in this case,
the splitting has almost no effect on the degree of entanglement, that
is maximum. Here, the leapfrog mechanism plays at its full extent: the
virtual states, on top of being degenerated and thus immune to the
splitting, remain always far, and are therefore protected, from the
real states. This results in the exactly constant phase (black panel
on the lower right end of Fig.~\ref{fig:6}). As a result, both
$C_\Gamma^\mathrm{int}(\tau_\mathrm{max})$ and $C_\Gamma(\tau)$ remain
large. The only drawback of this mechanism is, being virtual-processes
mediated, a comparatively weaker intensity.

\begin{figure}[t] 
  \includegraphics[width=\linewidth]{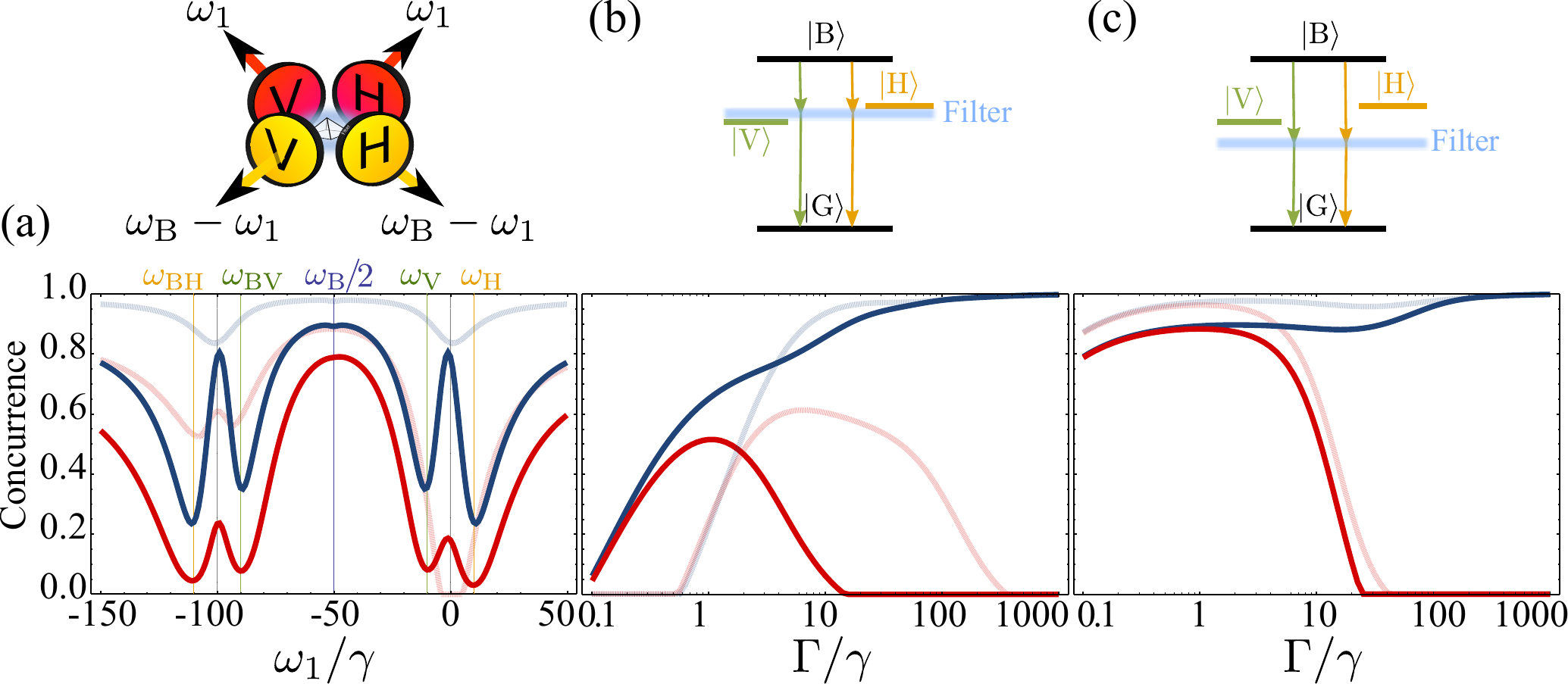}
  \caption{Concurrence computed with the instantaneous emission at
    $\tau=0$ (blue) or integrated up to $\tau_\mathrm{max}=1/\gamma$
    (red). (a) Plotted for a fixed filter linewidth $\Gamma=5\gamma$
    as a function of the filters frequencies
    $(\omega_1;\omega_\mathrm{B}-\omega_1)$. The concurrence is high
    unless a real state is probed and a cascade through it overtakes
    the leapfrog emission. The optimum case is the two-photon
    resonance. (b--c) Plotted as a function of the filter linewidth
    $\Gamma$ for the two cases that maximise entanglement in (a), that
    is at (b) the degenerate cascade configuration
    $(\omega_\mathrm{BX};\omega_\mathrm{X})$ and (c) the biexciton
    two-photon resonance $(\omega_\mathrm{B}/2;\omega_\mathrm{B}/2)$.
    The concurrence of the ideal case at $\delta=0$ is also plotted as
    a reference with softer solid lines. Parameters: $P=\gamma$,
    $\chi=100\gamma$, $\delta=20\gamma$.  $\omega_\mathrm{X}=0$ is set
    as the reference frequency.}
  \label{fig:7}
\end{figure}

We conclude with a more detailed analysis of what appears to be the
most suitable scheme to create a robust entanglement, the leapfrog
photon-pair emission. The target state is always $\ket{\psi}$,
Eq.~(\ref{eq:MonOct15142628CEST2012}), due to the degeneracy in the
filtered paths. The concurrence is shown as a function of the first
photon frequency, $\omega_1$, in Fig.~\ref{fig:7} (the second photon
has the energy $\omega_2=\omega_\mathrm{B}-\omega_1$ to conserve the
total biexciton energy). The time-integrated (resp.~instantaneous)
concurrence $C_\Gamma^\mathrm{int}(1/\gamma)$ [resp.~$C_\Gamma(0)$] is
shown in red (resp.~blue) lines, both for a large splitting
$\delta=20\gamma$ in strong tones and for $\delta=0$ in softer
tones. In the first panel, (a), the frequency $\omega_1$ of the
filters are varied. This figure shows that concurrence is very high
(with high state purity) when the filtering frequencies are far from
the system resonances: $\omega_\mathrm{BH/BV}$ and
$\omega_\mathrm{H/V}$. These are shown as coloured grid lines to guide
the eye. In this case, the real states are not involved and the
leapfrog emission is efficient in both polarisations. The concurrence
otherwise drops down when $\omega_1$ is resonant with any one-photon
transition, meaning that photons are then emitted in a cascade in one
of the polarisations, rather than simultaneously through a leapfrog
process.  Moreover, if at least one of the two deexcitation paths is
dominated by the real state dynamics, this brings back the problem of
``which path'' information, that spoils indistinguishability and
entanglement. There is only one exception to this general rule,
namely, the $\delta=0$ integrated case (soft red line) which has a
local maximum at $\omega_\mathrm{BX}$, i.e., when touching its
resonance in the natural cascade order. This is because the paths are
anyway identical and the integration includes the possibility of
emitting the second photon with some delay (up to
$\tau_\mathrm{max}$). For the case of $\delta\neq 0$, if this is large
enough, it is still possible to recover identical paths while
filtering the leapfrog in the middle points,
$\omega_1=\omega_\mathrm{BX}$ and $\omega_2=\omega_\mathrm{X}$, to
produce entangled pairs.

Overall, the optimum configuration is, therefore, indeed at the middle
point $\omega_1=\omega_2=\omega_\mathrm{B}/2$ (two-photon resonance),
where the photons are emitted simultaneously, with a high purity, and
entanglement degrees are also identical in frequency by construction.
Entanglement is also always larger in the simultaneous concurrence
(blue lines) as this is the natural choice to detect the leapfrog
emission, which is a fast process. This comes at the price,
expectedly, of decreasing the total number of useful counts and
increasing the randomness of the source. Note finally that the blue
curves are symmetric around $\omega_\mathrm{B}/2$, but the red ones
are not, given that in the integrated case, the order of the photons
with different frequencies is relevant. The left-hand side of the
plot, $\omega_1<\omega_\mathrm{B}/2$, corresponds to the natural order
of the frequencies in the cascade, $\omega_1<\omega_2$. The opposite
order, being counter-decay, is detrimental for entanglement.

Of all leapfrog configurations, those in panels (b) and (c) of
Fig.~\ref{fig:6} are therefore the optimum cases to obtain high
degrees of entanglement. Let us conclude with a study on their
dependence on the filter linewidths, $\Gamma$, in Figs.~\ref{fig:7}(b)
and~(c). First, we observe that in the limit of small linewidths for
the filters, i.e., in the region $\Gamma<1/\tau_\mathrm{max}$, the
simultaneous and time-integrated concurrences should converge to each
other, since the time resolution becomes larger than the integration
time. Therefore, the integrated emission provides the maximum
entanglement when the frequency window is small enough to provide the
same results than the simultaneous emission.  Decreasing $\Gamma$
below $P$ (which in this case is also $P=\gamma$) results in a time
resolution in the filtering larger than the pumping timescale
$1/P$. Therefore, photons from different pumping cycles start to get
mixed with each others. As a result $C$ drops for $\Gamma < \gamma$ in
all cases. In the limit $\Gamma \ll P$, the emission is completely
uncorrelated and $C=0$. This is an important difference from the cases
of pulsed excitation or the spontaneous decay of the system from the
biexciton state. In the absence of a steady state, entanglement is
maximised with the smallest window, $\Gamma\rightarrow
0$~\cite{akopian06a,meirom08a}. Two opposite behaviours are otherwise
observed in these figures when increasing the filter linewidth and
$\Gamma>1/\tau_\mathrm{max}$. The simultaneous emission gains in its
degree of entanglement whereas the time-integrated one loses with
increasing linewidth. In the limit $\Gamma\rightarrow\infty$, this
disparity is easy to understand. We recover the colour blind result in
all filtering configurations, (b) or (c), that is, the decay of
entanglement: from 1 corresponding to $\ket{\psi}$ at $\tau=0$, to 0
corresponding to a maximally mixed state at large $\tau$. Therefore,
$C_\Gamma(0)\rightarrow C(0) =1$ and
$C_\Gamma^\mathrm{int}(\tau_\mathrm{max})\rightarrow
C^\mathrm{int}(\tau_\mathrm{max}) = 0$ (for our particular choice of
$\tau_\mathrm{max}$ and $P$).

The decrease in $C_\Gamma^\mathrm{int}(\tau_\mathrm{max})$ when
increasing the filtering window, has been discussed in the literature
for the case in Fig.~\ref{fig:7}(b)~\cite{akopian06a,meirom08a}, and
it has been attributed to a gain of ``which path'' information due to
the overlap of the filters with the real excitonic levels. In the
lights of our results, when $\Gamma>\delta$ the real state
deexcitation takes over the leapfrog and
$C_\Gamma^\mathrm{int}(\tau_\mathrm{max})$ is suppressed indeed due to
such a gain of ``which path'' information, as discussed
previously. However, we find another reason why
$C_\Gamma^\mathrm{int}(\tau_\mathrm{max})$ decreases with $\Gamma$ in
all cases, based on the leapfrog emission: the region
$1/P<\Gamma<\delta$ for case (b) and the region $1/P<\Gamma<\chi$ for
case (c). The maximum delay in the emission of the second photon in a
leapfrog processes is related to $1/\Gamma$, due to its virtual
nature. Therefore, the initial enhancement of entanglement starts to
drop at delays $\tau\approx 1/\Gamma$, after which the emission of a
second photon is uncorrelated to the first one (not belonging to the
same leapfrog pair). For a fixed cutoff $\tau_\mathrm{max}$, this
leads to a reduction of $C_\Gamma^\mathrm{int}(\tau_\mathrm{max})$
with $\Gamma$. Broader filters have a smaller impact on the case of
real state deexcitation [see the zero splitting case in
Fig.~\ref{fig:7}(b), plotted in soft red]: since the system dynamics
is slower than the filtering ($\gamma$, $P<\Gamma$), the filter merely
emits the photons faster and faster after receiving them from the
system. This results in a mild reduction of
$C_\Gamma^\mathrm{int}(\tau_\mathrm{max})$ with $\Gamma$ until the
detection becomes colour blind and it drops to reach its aforementioned
limit of zero.

\section{Conclusions}
\label{sec:con}

In summary, we have characterised the emission of a quantum dot
modelled as a system able to accommodate two excitons of different
polarisation and bound as a biexciton. Beyond the usual single-photon
spectrum (or photoluminescence spectrum), we have presented for the
first time the two-photon spectrum of such a system, and discussed the
physical processes unravelled by frequency-resolved correlations and
how they shed light on various mechanisms useful for quantum
information processing.

We relied on the recently developed formalism~\cite{delvalle12a} that
allows to compute conveniently such correlations resolved both in time
and frequency. This describes both the application of external filters
before the detection or due to one or many cavity modes in
weak-coupling with the emitter. Filters and cavities have their
respective advantages, and when combined, can realize a distillation
of the emission, by successive filtering that enhance the correlations
and purity of the states.

We addressed three different regimes of operation depending on the
filtering scheme, namely as a source of single photons, a source of
two-photon states (both through cascaded photon pairs and simultaneous
photon pairs) and as a source of polarisation entangled photon pairs,
for which a form of the density matrix that is close to the
experimental tomographic procedure was proposed.  In particular,
so-called leapfrog processes---where the system undergoes a direct
transition from the biexciton to the ground state without passing by
the intermediate real states but jumping over them through a virtual
state---have been identified as key, both for two-photon emission and
for entangled photon-pair generation. In the latter case, this allows
to cancel the notoriously detrimental splitting between the real
exciton states that spoils entanglement through a ``which path''
information, since the intermediate virtual states have no energy
constrains and are always perfectly degenerate.  Entanglement is
long-lived and much more robust against this splitting than when
filtering at the system resonances.  At the two-photon resonance,
degrees of entanglement higher than $80\%$ can be achieved and
maintained for a wide range of parameters.

\section*{Acknowledgements}

The author acknowledges support from the Alexander von Humboldt
Foundation.

\section*{References}

\bibliographystyle{unsrt}
\bibliography{Sci,books,arXiv}
\end{document}